# Shaping liquid films by dielectrophoresis

Israel Gabay[1], Federico Paratore[2], Evgeniy Boyko[1,3], Antonio Ramos[4], Amir D.Gat[1*], Moran Bercovici[1*]

[1]*Faculty of Mechanical Engineering, Technion–Israel Institute of Technology, Haifa, Israel*

[2]*IBM Research Europe, Zurich, Switzerland*

[3]*Department of Mechanical and Aerospace Engineering, Princeton University, Princeton, USA*

[4]*Depto. Electrónica y Electromagnetismo, Facultad de Física, Universidad de Sevilla, Sevilla, Spain*

*corresponding authors: M.B. (mberco@technion.ac.il) and A.D.G (amirgat@technion.ac.il)

## ABSTRACT

We present a theoretical model and experimental demonstration of thin liquid film deformations due to a dielectric force distribution established by surface electrodes. We model the spatial electric field produced by a pair of parallel electrodes and use it to evaluate the stress on the interface through Maxwell stresses. By coupling this force with the Young-Laplace equation, we obtain the deformation of the interface. To validate our theory, we design an experimental setup which uses microfabricated electrodes to achieve spatial dielectrophoretic actuation of a thin liquid film, while providing measurements of microscale deformations through digital holographic microscopy. We characterize the deformation as a function of the electrode-pair geometry and film thickness, showing very good agreement with the model. Based on the insights from the characterization of the system, we pattern conductive lines of electrode pairs on the surface of a microfluidic chamber and demonstrate the ability to produce complex two-dimensional deformations. The films can remain in liquid form and be dynamically modulated between different configurations or polymerized to create solid structures with high surface quality.

## 1. Introduction

Dielectrophoresis (DEP) is a particular case of a force arising from the Maxwell stresses acting on dielectric materials containing permittivity gradients. The effect of DEP on particles has been studied extensively for over seven decades (Hughes, 2000; Jones, 1995; Pohl, 1978). Significant advancement in microfabrication techniques in the early '90 led to wider adoption of DEP, particularly in biological applications, as a method for control and manipulation of cells, viruses, proteins, and DNA (Chiou et al., 2005; Eberle et al., 2018; Hughes, 2000; Zhang et al., 2018).

To date, only a few studies have examined the effect of DEP forces in geometries which are not particles, bubbles, or droplets immersed in a liquid. Pellat (Pellat, 1894) was the first to study the effect of DEP on the rise of a dielectric liquid contained between two parallel electrodes. Extending Pellat's study, Jones et al. (Jones, 2002; Jones et al., 2003, 2004) investigated the influence of the liquid properties and the electric field frequency on the final height of the rising liquid, and used the term "liquid DEP" (credited to Melcher (Jones, 2002)) to describe problems associated with the motion of a liquid-air interface. Jones (Jones et al., 2001) also studied the use of DEP forces for the generation and control of micro- and nano-droplets on solid surfaces.

DEP actuation can also be leveraged for shaping the interface of a dielectric liquid film, as first demonstrated by Brown et al. (Brown et al., 2009, 2010, 2011; Wells et al., 2011). In their original work, Brown et al. demonstrated the ability to deform a thin liquid film using DEP forces produced by an array of interdigitated electrodes. Using electrode spacing between $20\,\mu m$ and $240\,\mu m$, they formed fluidic diffraction grating with corresponding peak-to-peak spacings (Brown et al., 2009). In follow-up work, they provided a detailed experimental study of the deformation scaling with the electric field magnitude, electrode spacing, and film thickness (Brown et al., 2010). Under the assumption of a periodic structure, and using a minimum energy approach, Brown et al. also provided an explicit analytical expression for the deformation (Brown et al., 2011). Wells et al. then showed that such structures could also be polymerized to yield solid optical components that no longer require the electric field to maintain their shape (Wells et al., 2011). Their demonstration naturally raises the desire to create other, more complex, surface deformations. However, such generic deformations cannot be addressed by a periodic energy-balance approach.

In this work, we provide an alternative approach for the analysis for thin film deformations by DEP forces that does not rely on the assumption of a periodic structures. Instead of energy approach, we utilize a force balance approach and model the spatial electric field created by pairs of electrodes patterned at the bottom of the fluidic chamber and calculate numerically the force distribution on the interface through Maxwell stresses. Coupling the DEP force with the Young-Laplace equation, we derive the governing equation describing the deformation of the interface. To validate the theory, we design an experimental setup which allows spatial dielectrophoretic actuation, measurements of the microscale deformations, and rapid curing of the deformed film into a solid object. Based on the insights from the characterization of the system, we demonstrate the ability to produce complex two-dimensional structures and provide guidelines for the design of such systems.

## 2. Concept and physical mechanism

Figure 1 presents the concept of thin liquid deformation using dielectrophoretic forces. The system we consider consists of a fluidic chamber filled with a thin layer of dielectric liquid rested on top of a rigid substrate containing patterned electrodes. Aiming to achieve localized deformations, we use pairs of closely spaced electrode lines to define arbitrary paths along the chamber, as illustrated in Figure 1A. Upon setting an AC electric potential difference between the electrodes, a strong localized electric field is created. Figure 1B presents the electric field lines at the high frequency regime where the liquid acts like a perfect dielectric (Castellanos et al., 2003; Morgan & Green, 2003). This electric field creates localized Maxwell stresses at the liquid-air interface (Figure 1C), which in turn act to deform the interface (Figure 1D). The permittivity difference in our system creates purely positive forces that pushes the interface upward yet, due to mass conservation, both positive and negative deformations are obtained. Figure 1E presents the deformation of a thin layer of silicone oil by DEP forces induced by an electrode pattern in the form of the word 'DEP'.

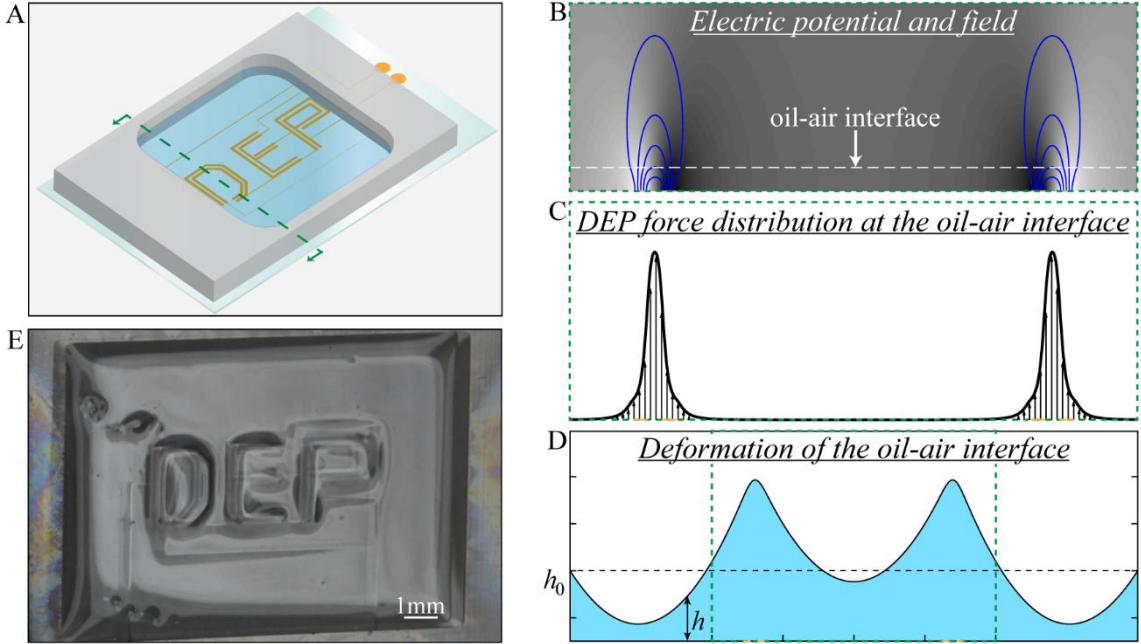

*Figure 1. Illustration of the concept of DEP-induced deformations. (A) Isometric view of the device used for inducing deformation, which consists of an open microfluidic chamber whose floor is patterned with pairs of electrodes leading to interface pads. The chamber is filled with a thin dielectric liquid film, forming a liquid-air interface. (B) A cut-view of the chamber showing that upon actuation of the electrodes, a non-uniform electric field is established (potential map indicated in grayscale, electric field lines in blue). (C) The electric field induces Maxwell stresses on the oil-air interface with maxima in proximity to the electrode pairs. (D) The stresses deform the liquid-air interface, with the deformation extending far beyond the electrodes region. While the DEP force is non-negative everywhere, mass conservation dictates both positive and negative deformation. The green dashed lines in D indicate the corresponding region shown in B and C. (E) Image of a rectangular microfluidic chamber filled with silicone oil and patterned with the same electrode configuration presented in A, where actuation of the electrodes enables to reshape the oil-air interface creating complex and localized pattern such as writing the word 'DEP'.*

## 3. Results
### 3.1. Theoretical model
Consider a liquid resting on top of a rigid substrate patterned with conductive electrodes and open to the air from the top, as illustrated in Figure 1A with a cartesian coordinate system at the bottom where the $x-y$ plane is parallel to the floor and the $z$ coordinate pointing at the normal direction to the floor. The electric body force at any point in the system can be expressed as (Melcher, 1981)

$$\mathbf{f} = \rho_E \mathbf{E} - \frac{1}{2} E^2 \nabla \varepsilon + \frac{1}{2} \nabla \left( \rho \frac{\partial \varepsilon}{\partial \rho} E^2 \right) \quad [1]$$

where $\rho_E$ is the free charge density, $\mathbf{E}$ is the electric field, $\varepsilon$ is the dielectric permittivity, and $\rho$ is the fluid's density. Under the assumptions of a high frequency regime, i.e., $\omega \gg \frac{\sigma}{\varepsilon}$ (where $\omega$ is the electric field frequency and $\sigma$ is the fluid's conductivity), the liquid can be assumed to behave as a perfect dielectric, and the contribution of the free charges (first term in [1]) vanishes (Castellanos et al., 2003; Ramos et al., 1998). Furthermore, if the liquid is assumed to have uniform permittivity, then the second term vanishes everywhere except at the interface where a discontinuity in permittivity exists. A convenient way to express the force distribution on the liquid-air interface is by considering the Maxwell stresses tensor, $\mathbf{T}_{ij}$, (Melcher, 1981) associated with [1],

$$\mathbf{T}_{ij} = \varepsilon \left( E_i E_j - \frac{1}{2} \delta_{ij} E_k E_k \left( 1 - \frac{\rho}{\varepsilon} \frac{\partial \varepsilon}{\partial \rho} \right) \right), \quad [2]$$

(where $\delta_{ij}$ is the Kronecker delta) and evaluating its normal projection on either side of this interface (see detailed derivation in SI section S1) provides the electrostatic force on the interface.

For convenience, we decompose the normal component of the electrostatic force on the interface, into two terms, representing the contribution of the permittivity discontinuity, $f_{DEP}$, and of the electrostriction, $f_{ES}$:

$$f_{DEP} = \frac{1}{2} \left( \varepsilon_f |\mathbf{E}_{a,\mathbf{t}}|^2 + \varepsilon_a E_{a,n}^2 \right) \left( 1 - \frac{\varepsilon_a}{\varepsilon_f} \right), \quad [3]$$

$$f_{ES} = \frac{1}{2} \left[ |\mathbf{E}_a|^2 \rho_a \left( \frac{\partial \varepsilon}{\partial \rho} \right)_a - |\mathbf{E}_f|^2 \rho_f \left( \frac{\partial \varepsilon}{\partial \rho} \right)_f \right]. \quad [4]$$

where $E_{a,n}$ and $\mathbf{E}_{a,\mathbf{t}}$ are the normal and tangential electric field components at the air side of the interface, respectively. The subscripts $a$ and $f$ mark the air and the fluid domains. We note that both the normal and tangential components of the electric field contribute to the force distribution in the normal direction to the interface.

To calculate the shape of the liquid-air interface at steady state, we express the normal stress balance while accounting for surface tension and Maxwell stresses,

$$P_f - P_a + f_{DEP} + f_{ES} = \gamma \kappa \quad [5]$$

where $\gamma$ and $\kappa$ are the surface tension and the mean-curvature of the liquid-air interface, $P_a$ is the pressure in the air (which can be assumed constant), and $P_f$ is the fluid pressure distribution on the interface. To resolve the pressure in the liquid, we write the electrohydrostatic equation, $\nabla p = \mathbf{f} - \rho_f \mathbf{g}$ (Stratton, 1941), which after integration yields

$$p_f(x,y,z) = P_0 + \rho_f \left( \frac{1}{2} \left( \frac{\partial \varepsilon}{\partial \rho} \right)_f |\mathbf{E}(x,y,z)|^2 - gz \right) \quad [6]$$

where $g$ is the gravitational acceleration, $p_f(x,y,z)$ is the pressure in the liquid, and $P_0$ is a constant that can be determined from the boundary conditions (e.g. far from the electrodes, at the liquid-air interface, $p_f$ is the atmospheric pressure). By substituting [4] and [6] into [5] we obtain

$$P_0 + \rho_f\left(\frac{1}{2}\left(\frac{\partial \varepsilon}{\partial \rho}\right)_f |\mathbf{E}|_f^2 - gh\right) - P_a + f_{DEP} + \frac{1}{2}\left[|\mathbf{E}_a|^2 \rho_a\left(\frac{\partial \varepsilon}{\partial \rho}\right)_a - |\mathbf{E}_f|^2 \rho_f\left(\frac{\partial \varepsilon}{\partial \rho}\right)_f\right] = \gamma \kappa, \quad [7]$$

where $h$ is the height of the liquid film. Importantly, the contribution of the electrostriction pressure in the fluid cancels out with the fluid's contribution to the electrostriction force distribution on the interface. Furthermore, using the Clausius-Mossotti law (Stratton, 1941),

$$\rho \frac{\partial \varepsilon}{\partial \rho} = \frac{(\varepsilon - \varepsilon_0)(\varepsilon + 2\varepsilon_0)}{3\varepsilon_0}, \quad [8]$$

the electrostriction pressure in the air vanishes ($\varepsilon_a = \varepsilon_0$). The final equation governing the shape of the interface is thus

$$f_{DEP} - \gamma \kappa - \rho g h = P_a - P_0. \quad [9]$$

### 3.1.1. Two-dimensional model

While the system we consider in our experimental work is three-dimensional, significant insight on the DEP forces and the deformations of the liquid-air interface can be obtained through analysis of a two-dimensional system. Consider a two-dimensional fluidic chamber of length $l$ and height $h_0$ filled with a dielectric liquid of volume $V_f$ (per unit depth) creating a thin liquid film, as illustrated in Figure 2. The floor of the chamber contains at its center a pair of electrodes of width and gap $l_e$ and negligible thickness. The dielectric permittivity of the liquid and the air above it are $\varepsilon_f$ and $\varepsilon_a$, respectively, and the surface tension of the liquid-air interface is $\gamma$.

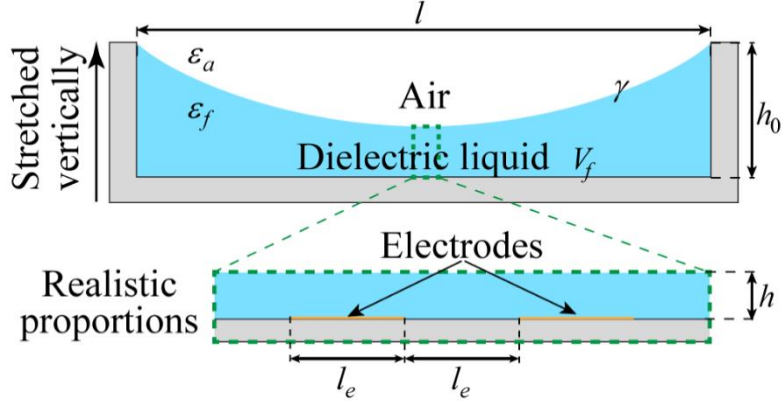

*Figure 2. Two-dimensional illustration of the parallel electrode pair configuration and the relevant physical parameters used in modelling the system. (A) A dielectric liquid of volume $V_f$ is placed in a chamber of length $l$ and height $h_0$, forming a thin film wetting the chamber's floor and walls. Two surface electrodes of width and gap $l_e$, are located at the center of the chamber. The dielectric permittivity of the fluid and air above it are $\varepsilon_f$ and $\varepsilon_a$, respectively, and the surface tension of the fluid-air interface is $\gamma$. (B) A closer view on the electrode region. Since the dimensions of the electrodes are significantly smaller than the size of the chamber, we assume an approximately constant height for the liquid film for the purpose of electric field and force calculations.*

The electric field and the deformation in the system are coupled. However, because the electric field decays rapidly away from the electrodes, we consider a simplified model in which the film thickness is uniform and equal to its value at the center between the electrodes, as illustrated in Figure 2B. This approximation holds well for small deformations, and as evident by the experimental results, provides very good predictions also for large deformations.

We numerically solve the electrostatic Laplace equation in a domain containing the two fluids. Substituting the resulting electric field into equation [3] yields the force distribution on the interface. Figure 3A presents the DEP force distribution (stress) at the liquid-air interface for electrodes widths $l_e = 120$ μm, $180$ μm, $240$ μm and a fixed film thickness of $h = 100$ μm, showing that the maximum is achieved midway between the electrodes and decreases as the electrodes are further gapped from one another. This is expected, because a larger distance between the electrodes leads to a proportionally smaller electric field.

Observing Figure 3A, one may mistakenly conclude that the total force per unit depth on the interface, $F_{DEP} = \int_{-l/2}^{l/2} f_{DEP} dx$, also decreases with the increase in $l_e$. However, as shown in the Figure 3D, the total force is non-monotonic, and the cases $l_e = 180$ μm, $240$ μm provide a larger total force than $l_e = 120$ μm, despite a lower maximum. Figure 3B presents the total force as a function of both $l_e$ and the film thickness $h$. Here, too, the non-monotonicity $l_e$ is evident across all $h$ values, where for any film thickness, the maximum total force is obtained when setting $l_e = 1.85h$. As a result, while for $h > 0.54l_e$ the total force increases with $l_e$, for $h < 0.54l_e$ the dependence is inverted and the total force decreases with $l_e$. This behaviour can also be seen more explicitly in Figure 3C, presenting the total force as a function of $h$ for different fixed $l_e$ values, equivalent to tracing Figure 3B along vertical lines. The different decay rates of the force with $h$, result in the curves intersecting one another.

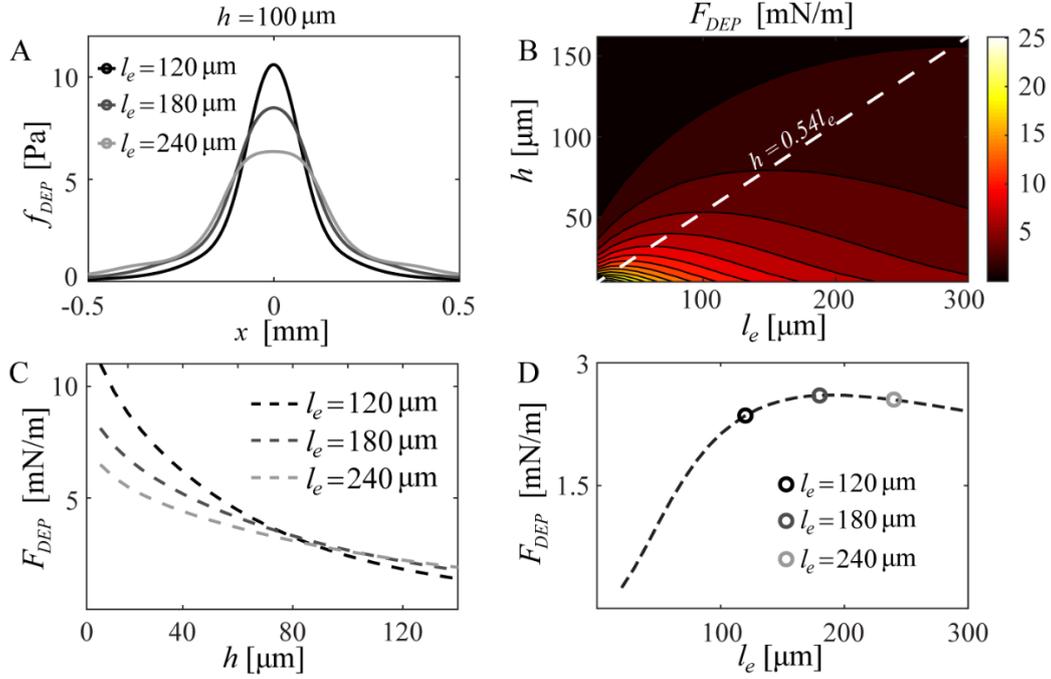

**Figure 3.** *Two-dimensional finite-element simulation results showing the behavior of the DEP force acting on the interface for electrodes pair configurations where the width of the electrodes is much smaller than the chamber length, $l_e \ll l$. (A) DEP force distribution on the interface along the chamber for three different electrode widths, with $h = 100$ μm, showing that the maximum achieved midway between the electrodes, decreases as the width of the electrodes increases. (B) A color map showing the total force on the oil-air interface, (integral over the DEP force distribution), as a function of the $l_e$ and the film thickness $h$. The white dashed line indicates the electrode width that provides that maximum force for a given $h$. (C) As expected, for a fixed electrode width, the force decreases as the liquid thickness increases. We note the cross-over point indicating that the dependence of the force on the electrodes gap is inverted for sufficiently large film thicknesses. (D) The total DEP force as a function of the electrodes' length for a fixed film thickness ($h = 100$ μm, showing a non-monotonic dependence. The circles correspond to each of the cases in subfigure A. The simulations were performed using $l = 9$ mm, $\varepsilon_f = 2.5\varepsilon_0$, $\varepsilon_a = \varepsilon_0$, and $V_0 = 400$ V.*

As presented in Figure 3A, the numerically obtained DEP force distributions strongly resemble a Gaussian. To facilitate an explicit expression for the deformation, we thus approximate $f_{DEP}$ as a Gaussian of width (standard deviation) $l_e$, and an amplitude $a$ set such that its total force matches $F_{DEP}$,

$$f_{DEP} = a\exp\left[-\left(\frac{x}{l_e}\right)^2\right], \quad a = \frac{F_{DEP}}{l_e\sqrt{\pi}\,\text{erf}\left[l/2l_e\right]}. \qquad [10]$$

Under the long-wave approximation (Leal, 2007; Oron et al., 1997), the interface curvature can be approximated as $\frac{d^2h}{dx^2} = h''(x)$ and equation [9] can be written as

$$\gamma h'' - \rho g h + f_{DEP} = P_a - P_0, \qquad [11]$$

We assume that the liquid is pinned at the edges of the chamber, $h(-l/2) = h(l/2) = h_0$, providing two boundary conditions which are sufficient to solve equation [11] as a function of the constant pressure difference $P_a - P_0$. To then obtain the value of Pa-P0, we require the total fluid volume to be conserved, $\int_{-l/2}^{l/2} h(x)\,dx = V_f$. The complete solution, including the effect of gravity, is presented in the SI. For brevity, we here provide the more compact expression for the case of $g = 0$ and $V_f = h_0 l$. It is also convenient to present the solution in terms of the deformation of the liquid-air interface relative to its initial state, $d = h - h_0$,

$$d(\xi) = A\left(\left(-8e^{-\frac{\xi^2}{c^2}} + 2e^{-\frac{1}{c^2}}(1+3\xi^2)\right)c + \sqrt{\pi}\left((2+6\xi^2 - 3c^2(\xi^2 - 1))\text{erf}\left[\frac{1}{c}\right] - 8\xi\,\text{erf}\left[\frac{\xi}{c}\right]\right)\right) \qquad [12]$$

where $\xi = \frac{2x}{l}$ is the non-dimensional axial coordinate, $c = \frac{2l_e}{l}$ is the non-dimensional electrode width, and $A = \frac{F_{DEP} l}{16\gamma\sqrt{\pi}\,\text{erf}[1/c]}$.

In this work, we investigate only the case of $c \ll 1$, corresponding to narrow electrodes compared to the length of the chamber. At this limit, the deformation magnitude scales with A, and is linear with the chamber's length, $l$ and inversely proportional to the surface tension. Consistently with the assumption of $c \ll 1$, the spatial gradients of the deformation, $\frac{d}{dx}(d) = \frac{d}{d\xi}(d)\frac{2}{l}$, are independent of the chamber size $l$.

The solution for the deformation [12] is a function of the DEP force, which in turn depends on the fluid thickness. Therefore, to solve for the deformation for a given voltage we use an iterative solver. Using the DEP force calculated based on the initial fluid thickness, we obtain an initial solution for the deformation. Updating the DEP force using the obtained height of the interface above the electrodes yields a new solution for the deformation. We repeat the process until the relative change in interface height between iterations (evaluated at $x = 0$) reduces below $10^{-5}$.

## 3.2. Experimental measurements

In Figure 4A we present 3D digital holographic measurements (Cuche et al., 1999) of the deformation of an oil-air interface due to an electric field produced by a pair of parallel electrodes. The fluidic chamber has a width and length of $l = 9$ mm and depth of approximately $h_0 = 120$ μm. The electrodes have a width and gap of $l_e = 120$ μm and span the entire length of the chamber. The dielectric fluid is a low-viscosity silicone oil with density $\rho = 913$ kg/m$^3$, and surface tension $\gamma = 20$ mN/m. We initially measure the oil-air interface in the absence of an electric field. This measurement is subtracted from all subsequent measurements, thus allowing to isolate the deformation of the interface due to the electric field.

In panels 4B and 4C we compare the experimental data with the theoretical predictions for the deformation along the $x$-axis, for different electrode widths and different fluid thicknesses. Increasing the applied voltage increases the electric field and results in larger deformations. Similarly, decreasing the initial liquid volume, which decreases the distance of the interface from the electrodes, also results in a higher electric field and larger deformations. Both effects, as well as the details of the spatial deformation, are captured by the model, in very good agreement with the experimental results. The results show that the use of the long-wave approximation is well justified even for the largest deformations, where the maximal slope of the interface is on the order of $100$ μm over $1$ mm.

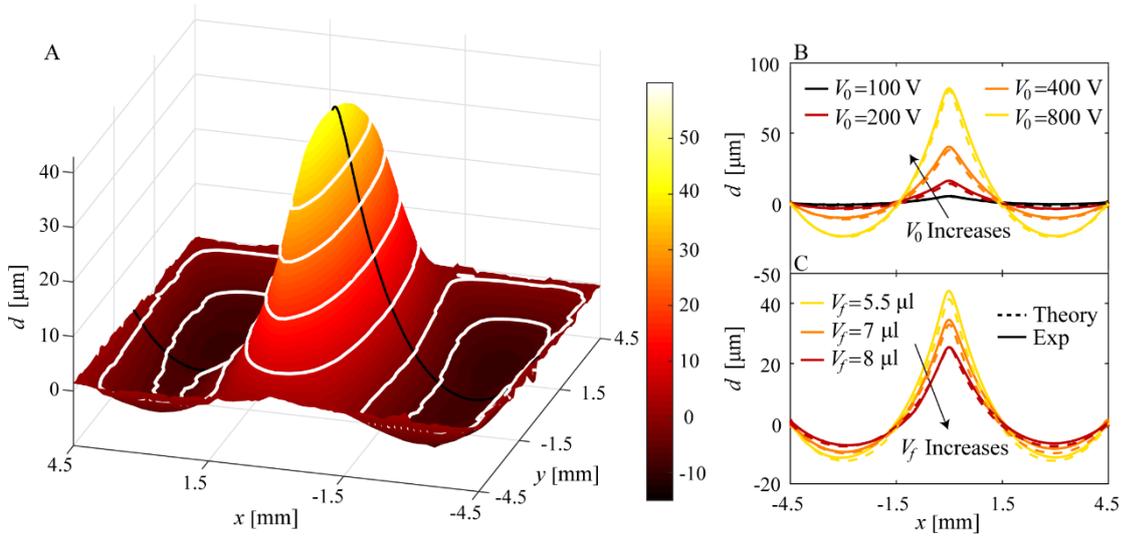

*Figure 4. Experimental measurements and theoretical predictions of DEP-induced deformations using the parallel electrodes pair configuration. (A) A typical experimental result showing the three-dimensional shape of the deformations, resulting from actuation of a pair of electrodes positioned along the $y$-axis. (B, C) The deformation along the $x$-axis at $y = 0$ for different applied voltages and initial liquid volumes, respectively. The solid lines present the experimental results, and the dashed lines present the theoretical predictions, obtained from the two-dimensional model Equation [11]. The maximum deformation is achieved in the middle of the chamber, between the electrodes, and it increases as the applied voltage increases and decreases as the fluid volume (fluid height above the electrodes) increases. We use silicone oil with a dielectric permittivity of $\varepsilon_f = 2.5\varepsilon_0$ and surface tension of $\gamma = 20$ mN/m, a square-shaped chamber with $l = 9$ mm, $h_0 = 120$ μm, $l_e = 120$ μm, and an AC voltage with a frequency of $10$ kHz.*

Figure 5A presents the maximum deformation as a function of the applied voltage square, showing good agreement between theory and experiments. At low voltages, the deformation magnitude follows well the $V^2$-dependence obtained from scaling of the DEP force. However, at higher voltages, the maximum deformations are lower than those suggested by the scaling. This is precisely because of the earlier mentioned coupling between the deformation and the electric force; at high voltages, the deformation becomes significant enough to affect (reduce) the electric field at the interface.

This coupling is also evident in Figure 5B that presents the predicted and measured (normalized) maximum deformations as a function of $l_e$, for different voltages. The dependence on $l_e$ is different for different voltages, to the extent that the trend is inverted between the lowest and highest voltages shown. This could be explained by the fact that higher voltages are associated with larger film thickness. For example, the case of $100\,\text{V}$ yields deformations of several microns, whereas the $800\,\text{V}$ case yields deformations of approximately $100\,\mu\text{m}$ (both relative to an initial thickness of $35\,\mu\text{m}$). These film thickness values reside on opposite sides of the intersection region shown in Figure 3C, and thus the inverted dependence on $l_e$ is expected. This result elucidates that to achieve the maximum force, the electrodes width and gap $l_e$ should be chosen such that $l_e = 1.85h$ in accordance with Figure 3B, but where $h$ is the post-deformation film thickness, rather than the initial one.

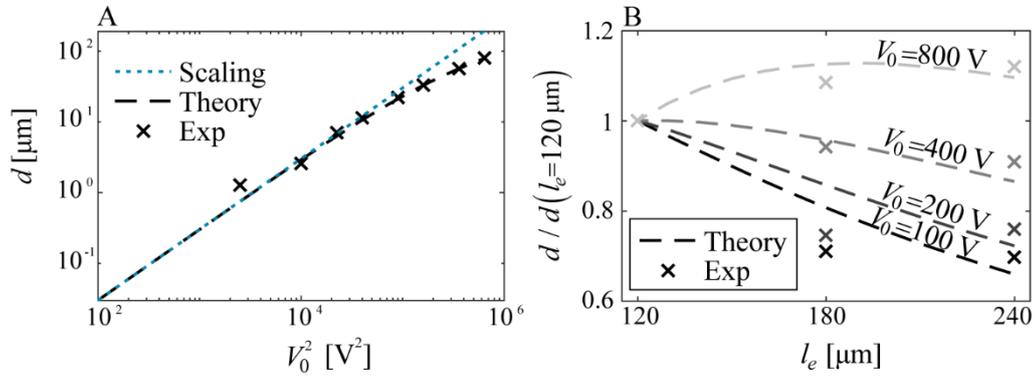

*Figure 5.* Comparison of experimental and theoretical results of the maximum deformation. (A) The maximal deformation as a function of the voltage squared for $l_e = 120\,\mu\text{m}$ and $V_f = 8\,\mu\text{l}$. The black dashed line presents the theoretical prediction based on the one-dimensional model, the black crosses present the experimental results, and the blue dashed line represents the linear scaling with $V_0^2$. When the deformation is small compared to the initial fluid thickness above the electrodes, the theoretical solution scales linearly with $V_0^2$, yet as the deformation becomes comparable to the initial film thickness, both theoretical and experimental results show a sub-linear behaviour with $V_0^2$, due to the inverse scaling of the DEP force with the film thickness. (B) Normalized maximum deformation as a function of the electrodes width for different voltages. The dashed lines present the theoretical predictions, and the crosses represent the experimental results. For low voltages (e.g., $V_0 = 100\,\text{V}$, black line), the deformation decreases when the electrodes width $l_e$ increases, but above a certain value of $V_0$ the deformation increases as $l_e$ increases (e.g., $V_0 = 800\,\text{V}$, light gray line). This transition is associated with the cross-over in the total force $F_{DEP}$ observed in Figure 3C.

### 3.3. Fluid shaping

The electrode pair configuration can serve as a basic unit for the creation of complex two-dimensional electrode structures. Figure 6A.1 presents experimental measurements of the oil-air interface topography resulting from the actuation of an electrode configuration tracing the letters 'DEP' on the surface (Figure 6B.1). The deformation clearly shows peaks along the electrode pairs, and the letters are clearly distinguishable. However, as also visible from the cross-section in Figure 6C.1, the peaks are not well separated. During our experiments, we found that reducing the liquid volume can significantly increase the resolution and contrast of the deformation field. Figure 6B.2 presents the same electrode configuration and applied voltage, but with $2\,\mu\text{l}$ instead of $4\,\mu\text{l}$ of liquid, resulting in its accumulation primarily at the edges of the chamber and only minimally wetting of the floor. As a result, upon actuation of the voltage, the liquid is drawn from the edges of the chamber toward the electrodes. Since the proximity of the floor precludes significant negative deformations, as can be seen clearly from both Figs. 6A.2 and 6C.2, the resulting deformation shows a better separation of the peaks and consequently better-defined letters. The same conditions can be applied to other electrode configurations, as shown in Figure 6.3. Here, the electrode pairs are patterned to form an outline of a Y-junction. Upon activation of the electric field, each electrode pair produces a vertical wall, forming the physical boundaries of a $60\,\mu\text{m}$ deep and $1\,\text{mm}$ wide Y-junction channel. As shown in Movie S1, upon activation of the field, the deformation is rapidly formed, can be easily modulated in amplitude, turned on and off, and quickly recovers from external forced disturbances.

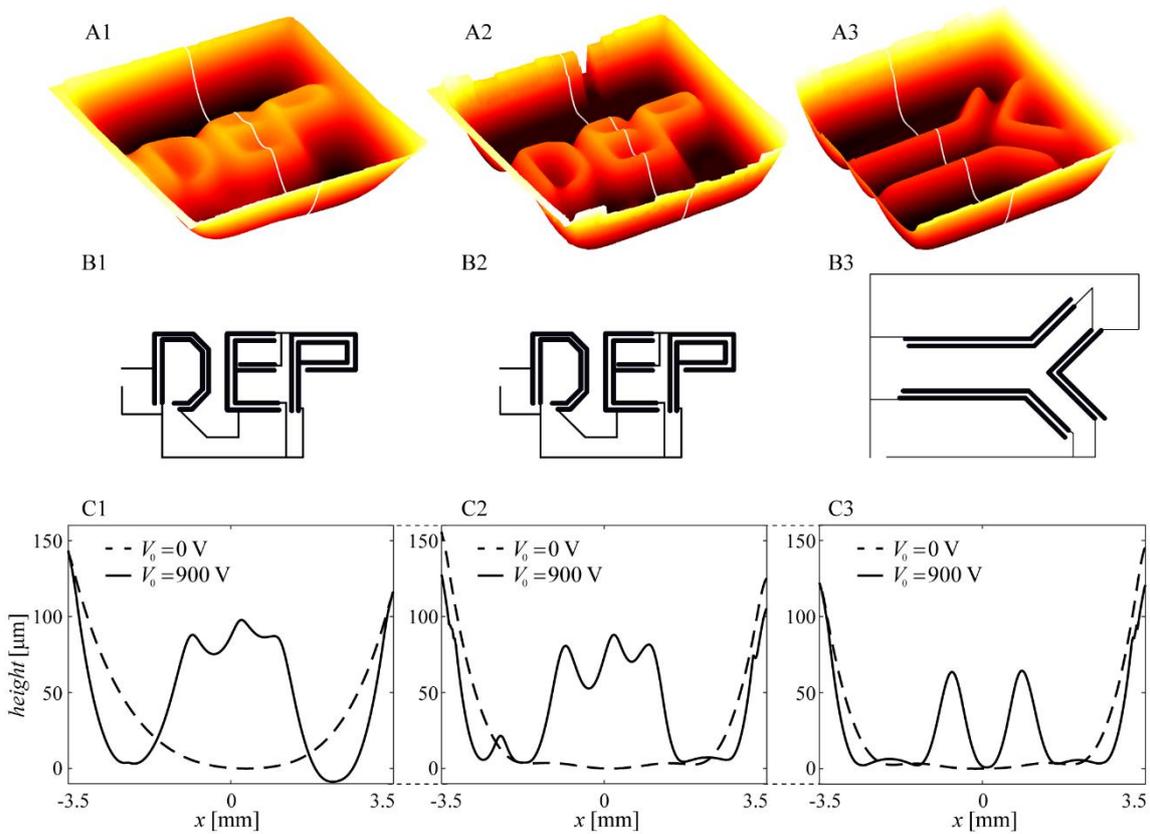

**Figure 6.** Experimental results demonstrating the use of DEP-based deformation for the creation of complex structures. Each configuration is based on pairs of electrodes deposited on a desired pattern at the bottom of the fluidic chamber (B). Upon actuation of the electric field, the liquid deforms to obtain the desired shape corresponding to the electrode configuration. The dashed and solid black curves in the two-dimensional images present the shape of the interface before and after actuation, respectively, along the $x$-axis denoted by a white line in the three-dimensional figure. (1) Using $4\,\mu\text{l}$ of liquid, the initial interface is curved and the displacement of the liquid from the periphery into the actuation region is distinctly visible in C1. A1 shows the resulting topography which reads 'DEP'. (2) using only $2\,\mu\text{l}$ of liquid, the initial interface at the center of the chamber is nearly flat, which results in accentuation of the deformations and improved resolution relative to the $4\,\mu\text{l}$ case, providing

better contrast and readability. (3) Using $2\,\mu l$, we demonstrate the creation of a $\sim 1\,\text{mm}$ wide, $\sim 60\,\mu\text{m}$ deep microfluidic channel and a Y-junction.

Due to the nature of the liquid-air interfaces, the resulting surfaces of the produced structures are very smooth. Thus, replacing the silicone oil with a polymer opens the door to a fabrication of smooth solid structures. Figure 7A presents an example for a structure produced by deforming the interface of a photopolymer using the Y-junction electrode configuration; after the steady-state deformation is obtained, and while the electrodes are still active, we expose the film to $365\,\text{nm}$ UV light for five minutes which leads to its solidification. Figure 7B presents the shape of the interface along a cross-section before polymerization (i.e., in liquid state) and following polymerization (i.e., in solid state), showing good agreement between the two. Figure 7C presents the surface quality of the solidified part. Fitting the measured data to a second-degree polynomial and subtracting it from the original curve provides an estimate for the surface roughness, which is on the order of $3.2\,\text{nm}$ root mean square (RMS) (Figure 7D). The surface roughness may in fact be better, as this is the limitation of the digital holographic microscope, as shown by measurements of an atomically polished wafer (SI Figure S5).

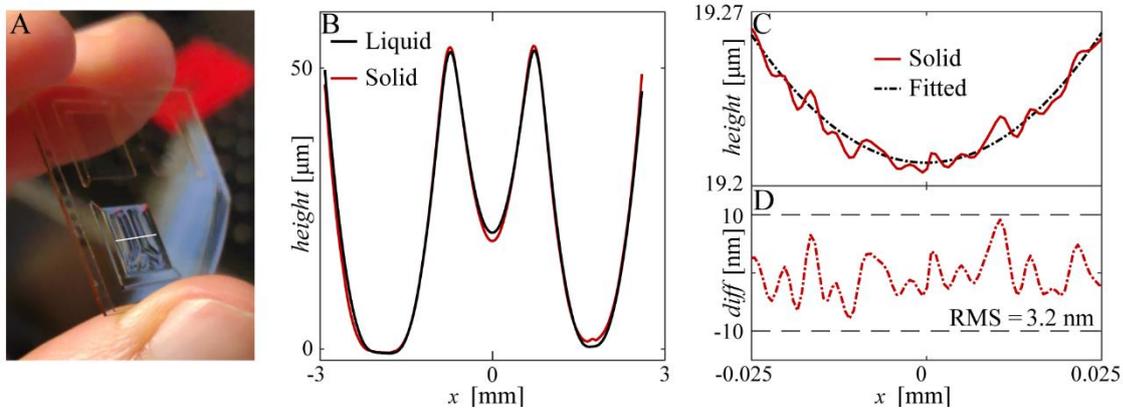

**Figure 7.** Experimental demonstration of the use of DEP-based deformation for the fabrication of smooth solid structures. (A) Image of a Y-junction fabricated by deformation and polymerization of a photopolymer. (B) Comparison of the cross-section along the chamber (indicated by the white dashed line in A) before and after solidification of the polymer. (C) Zoomed-in view of the surface at $x=0$ (solid red curve) together with a 2nd order polynomial fit (black dashed curve). (D) The difference between the raw data and the fit provides an estimation of the surface quality, yielding an RMS value of $3.2\,\text{nm}$.

## 4. Methods

We fabricated the devices using standard cleanroom microfabrication processes. We used a 4-inch borosilicate glass wafer (Borofloat33, Wafer Universe, Germany) as a substrate on which we patterned *via* lift-off a $6\,\text{nm}$ layer of an of 2 nm titanium – 2 nm platinum – 2 nm titanium. We used such a thin layer because it is semi-transparent in the visible spectrum and thus reduces the reflection of the holographic microscope's laser beam. We defined the fluidic chamber's walls by lithography processing of a $120-150\,\mu\text{m}$ thick layer of SU8, created by spin-coating of SU8-50 (Microchem AG, Germany) in two sequential steps.

We performed the experiments by placing at the center of the chamber few microliters of low-viscosity silicone oil (Cat. No. 317667, Sigma-Aldrich) of density $\rho = 913\,\text{kg/m}^3$, refractive index $n = 1.403$, and surface tension $\gamma = 20\,\text{mN/m}$, measured using an optical tensiometer (Theta Flex, Biolin Scientific). We used a wave generator (TG5012A, AIM-TTI Instruments) connected to an amplifier (2210-CE, TREK) to deliver to the electrodes a 10 kHz sinusoidal AC electric potential at voltages (peak to peak) of up to $900\,\text{V}$.

The measurements of the DEP induced deformations were obtained using a digital holographic microscope (DHM-R1003, Lyncee Tec) through a 10X objective with a field of view of $0.5 \text{X} 0.5 \text{ mm}^2$ (see detailed explanation in SI section 3). To obtain full coverage of the chamber area we used an automated stage (MS 2000, ASI) working in synchronization with the DHM camera. We dictated a constant movement of $407\,\mu\text{m}$ and stitched the data to assemble the image of the entire oil-air interface.

For the polymerization experiments, we used a UV curable polymer (CPS 1050, Colorado Polymer Solutions) and activated the electrodes at a voltage of 400 V and frequency of 10 kHz. We solidified the polymer using two $12\,\text{W}$ UV lamps with a wavelength of $365\,\text{nm}$ for 5 minutes.

## 5. Discussion and Conclusions

We presented a theoretical model and an experimental demonstration of a new and practical approach to create desired deformations of a liquid-fluid interface. Owing to their inherently smooth interfaces, the ability to shape liquid films holds great promise as a method to create and modulate optical components. We showed that the use of pairs of electrodes provides an effective method for creating desired deformations. Beyond their ability to form highly localized deformations, from a practical perspective, continuous parallel electrodes allow to span significant portions of the working area using only a single connection at the edge of each electrode. Furthermore, we showed that the distance between the electrodes can be used to control the magnitude of deformation, allowing the deformation to vary along the electrode-pair path.

In the current work, we studied the steady-state deformation of the system. When using a polymer, the liquid film can be solidified to yield a permanent component that could be used outside of the DEP system. However, one potential advantage of a fluidic system, particularly in the context of adaptive optics, is the ability to dynamically modulate it, transitioning from one configuration to another. Movie S2 demonstrates this concept using an array of parallel electrodes, where the actuation transitions dynamically between one set to another.

We focused on deformations of an oil-air interface, with a dielectric constant ratio of approximately 2.5. As indicated by equation [3], the force is proportional to this ratio, and thus much larger deformations can be expected when using liquids with a higher dielectric constant. A natural candidate would be water, with a relative permittivity that is ~30-fold greater than that of silicone oil. For sufficiently high frequencies, such a system can be considered to be governed by dielectric effects, yet for lower frequencies, one must consider conductivity effects that would not only alter the force on the interface (see SI section 1) but would also lead to additional effects such as Joule heating and internal flows which we did not consider in this work. Our theory can also be directly applied to liquid-liquid configurations, providing an opportunity to invert the permittivity ratio relative to the oil-air configuration, i.e., have the liquid with the lower permittivity be in contact with the electrodes. In such a case, the resulting force on the interface will be toward the electrodes rather than away from them. This may lead to larger deformations, due to the pulling force further increasing as the interface approaches the surface. Beyond a certain threshold, this is also expected to lead to instability and rupture of the film over the electrode's region.

# Supplementary Information

# Shaping liquid films by dielectrophoresis

Israel Gabay[1], Federico Paratore[2], Evgeniy Boyko[1,3], Antonio Ramos[4], Amir D.Gat[1*], Moran Bercovici[1*]

[1]*Faculty of Mechanical Engineering, Technion–Israel Institute of Technology, Haifa, Israel*

[2]*IBM Research Europe, Zurich, Switzerland*

[3]*Department of Mechanical and Aerospace Engineering, Princeton University, Princeton, USA*

[4]*Depto. Electrónica y Electromagnetismo, Facultad de Física, Universidad de Sevilla, Sevilla, Spain*

*\*corresponding authors: M.B. ([mberco@technion.ac.il](mberco@technion.ac.il)) and A.D.G ([amirgat@technion.ac.il](amirgat@technion.ac.il))*

1. **Electric force distribution on the interface**

Consider two immiscible fluids with dielectric permittivity $\varepsilon_1$ and $\varepsilon_2$ separated by an interface with a surface charge density $\sigma_E$, as illustrated in Fig. S1., We define $\hat{\mathbf{t}}_1$ and $\hat{\mathbf{t}}_2$ as the tangential unit vectors and $\hat{\mathbf{n}}$ as the normal unit vector to the interface pointing from the lower to the upper fluid.

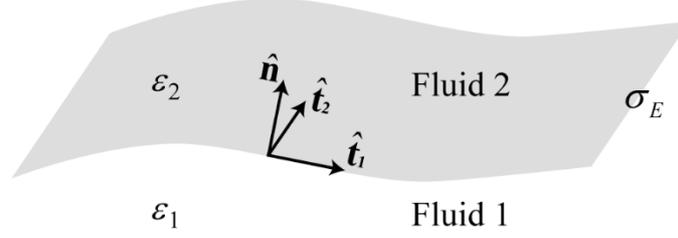

*Fig. S1. An illustration of an interface (gray surface) separating two immiscible fluids. The lower (fluid 1) and the upper (fluid 2) fluids have electric permittivity of $\varepsilon_1$ and $\varepsilon_2$, respectively, and $\sigma_E$ is the surface charge density at the interface.*

To derive the electric stress acting at the interface, we start from the Maxwell stress tensor (Melcher, 1981; Stratton, 1941),

$$\mathbf{T}_{ij} = \varepsilon \left( E_i E_j - \frac{1}{2}\delta_{ij} E_k E_k \left(1 - \frac{\rho}{\varepsilon}\frac{\partial \varepsilon}{\partial \rho}\right)\right), \quad [S1.1]$$

where $\varepsilon$ is the fluid permittivity, $E_i$ is the $i$ component of the electric field at each point, and $\delta_{ij}$ is the Kronecker delta. The Maxwell stress tensor includes the Coulombic force contribution (the force on free charges), the dielectric force contribution due to permittivity gradients, and the electrostriction force contribution. The electrical stress (force distribution) acting on the interface can be written as $\mathbf{f} = (\mathbf{T}_2 - \mathbf{T}_1)\cdot\hat{\mathbf{n}}$, where $\mathbf{T}_1$ and $\mathbf{T}_2$ are the Maxwell stress tensor contributions from the lower and the upper sides of the interface, respectively.

The explicit expression for the force distribution at the interface is therefore,

$$\mathbf{f} = \frac{1}{2}\left[\varepsilon_2\left(E_{2,n}^2 - |\mathbf{E}_{2,\mathbf{t}}|^2\right) + \varepsilon_1\left(|\mathbf{E}_{1,\mathbf{t}}|^2 - E_{1,n}^2\right) + |\mathbf{E}_2|^2 \frac{\rho_2}{\varepsilon_2}\left(\frac{\partial \varepsilon}{\partial \rho}\right)_2 - |\mathbf{E}_1|^2 \frac{\rho_1}{\varepsilon_1}\left(\frac{\partial \varepsilon}{\partial \rho}\right)_1\right]\hat{\mathbf{n}} + \varepsilon_2 \mathbf{E}_{2,\mathbf{t}} E_{2,n} - \varepsilon_1 \mathbf{E}_{1,\mathbf{t}} E_{1,n}$$
,[S1.2]

where $\mathbf{E}_1$ and $\mathbf{E}_2$ are the electric fields at the lower and the upper sides of the interface, respectively, $\mathbf{E}_{1,\mathbf{t}} \equiv \mathbf{E}_1 - E_{1,n}\hat{\mathbf{n}}$ and $\mathbf{E}_{2,\mathbf{t}} \equiv \mathbf{E}_2 - E_{2,n}\hat{\mathbf{n}}$ are the tangential components of the electric field, and $E_{1,n}$ and $E_{2,n}$ are the normal components of the electric field. We note that in the manuscript we marked the bottom fluid with subscript $f$ for fluid, and the upper fluid with the subscript $a$ for the air, here we kept them fluid 1 and fluid 2 to present the more general case of two general fluids.

Using the relations of the electric field components at the interface, the continuity of the tangential component of the electric field $\mathbf{E_{1,t}} = \mathbf{E_{2,t}}$, and the jump condition relating the normal components of the displacement field to the surface charge distribution at the interface, $\varepsilon_2 E_{2,n} - \varepsilon_1 E_{1,n} = \sigma_E$, we can express the tangential component of the electric force distribution at the interface as,

$$\mathbf{f} \cdot \begin{pmatrix} \mathbf{t_1} \\ \mathbf{t_2} \end{pmatrix} = \varepsilon_2 \mathbf{E_{2,t}} E_{2,n} - \varepsilon_1 \mathbf{E_{1,t}} E_{1,n} = \mathbf{E_{1,t}} \left( \varepsilon_2 E_{2,n} - \varepsilon_1 E_{1,n} \right) = \mathbf{E_{1,t}} \sigma_E = \mathbf{E_t} \sigma_E, \quad [S1.3]$$

where $\mathbf{E_t} \equiv \mathbf{E_{1,t}} = \mathbf{E_{2,t}}$.

For the normal component of the force distribution at the interface, we use the same relations for the normal and tangential electric field, and after some algebraic manipulations we obtain,

$$\mathbf{f} \cdot \hat{\mathbf{n}} = \frac{1}{2} \left[ \varepsilon_2 \left( E_{2,n}^2 - |\mathbf{E_{2,t}}|^2 \right) + \varepsilon_1 \left( |\mathbf{E_{1,t}}|^2 - E_{1,n}^2 \right) + |\mathbf{E_2}|^2 \rho_2 \left( \frac{\partial \varepsilon}{\partial \rho} \right)_2 - |\mathbf{E_1}|^2 \rho_1 \left( \frac{\partial \varepsilon}{\partial \rho} \right)_1 \right] =$$
$$= \frac{1}{2} \left[ \left( \varepsilon_1 |\mathbf{E_t}|^2 + \varepsilon_2 E_{2,n}^2 \right) \left( 1 - \frac{\varepsilon_2}{\varepsilon_1} \right) - \frac{\sigma_E^2}{\varepsilon_1} + \frac{2\varepsilon_2 E_{2,n} \sigma_E}{\varepsilon_1} + |\mathbf{E_2}|^2 \rho_2 \left( \frac{\partial \varepsilon}{\partial \rho} \right)_2 - |\mathbf{E_1}|^2 \rho_1 \left( \frac{\partial \varepsilon}{\partial \rho} \right)_1 \right]. \quad [S1.4]$$

Under the assumptions of an alternating current electric field at high frequency, such that the liquid acts as a dielectric, and isothermal conditions, implying that the liquid properties are uniform, there are not accumulation of free charges in the system, and particularly at the interface, i.e., $\sigma_E = 0$. Therefore, there are no Coulombic force and the force distribution at the interface arises solely from the discontinuity in the permittivity and fluid's density at the interface,

$$\mathbf{f} = \frac{1}{2} \left[ \left( \varepsilon_1 |\mathbf{E_t}|^2 + \varepsilon_2 E_{2,n}^2 \right) \left( 1 - \frac{\varepsilon_2}{\varepsilon_1} \right) + |\mathbf{E_2}|^2 \rho_2 \left( \frac{\partial \varepsilon}{\partial \rho} \right)_2 - |\mathbf{E_1}|^2 \rho_1 \left( \frac{\partial \varepsilon}{\partial \rho} \right)_1 \right] \hat{\mathbf{n}}. \quad [S1.5]$$

Equation [S1.5] clearly shows that both the normal and tangential components of the electric field contribute to the normal component of the electrostatic force, although the force is only in the normal direction to the interface.

For convenience, we decompose the normal component of the force distribution at the interface into two terms. The first term is the dielectric force contribution, $f_{DEP}$,

$$f_{DEP} = \frac{1}{2} \left[ \left( \varepsilon_1 |\mathbf{E_t}|^2 + \varepsilon_2 E_{2,n}^2 \right) \left( 1 - \frac{\varepsilon_2}{\varepsilon_1} \right) \right], \quad [S1.6]$$

and the second term is the electrostriction contribution $f_{ES}$,

$$f_{ES} = \frac{1}{2} \left[ |\mathbf{E_2}|^2 \rho_2 \left( \frac{\partial \varepsilon}{\partial \rho} \right)_2 - |\mathbf{E_1}|^2 \rho_1 \left( \frac{\partial \varepsilon}{\partial \rho} \right)_1 \right]. \quad [S1.7]$$

In the main text, we use the expression for $f_{DEP}$ and $f_{ES}$ when writing the normal stress balance for calculating the shape of the liquid-air interface at a steady state.

## 2. Calculating the shape of the interface

Consider a two-dimensional fluidic chamber of length $l$ and depth $h_0$ filled with a dielectric liquid with volume $V_f$, as illustrated in Fig. S2. The floor of the chamber contains at its center a pair of electrodes of width and gap $l_e$ and negligible thickness (Fig. S2B). The dielectric permittivity of the fluid and of the air above it are $\varepsilon_f$ and $\varepsilon_a$, respectively, and the surface tension of the fluid-air interface is $\gamma$. Assuming the fluid have a uniform permittivity, we modify the Young-Laplace equation (normal stress balance at the interface) to account for the DEP force distribution on the interface, and present the equation for the fluid's height in the chamber,

$$\frac{\gamma h''}{\left(1+h'^2\right)^{3/2}} - \rho_f g h + f_{DEP} = P, \qquad [\text{S2.1}]$$

where $h$ is the fluid's height, $g$ is the gravitational acceleration, $\rho_f$ is the density of the liquid, $P$ is a constant representing the pressure and can be determined from the boundary conditions, and primes represent differentiation with respect to $x$. Under the long-wave approximation (Leal, 2007; Oron et al., 1997), i.e., $\tilde{h}'^2 \ll 1$, we obtain the following linear equation for the shape of the interface,

$$\gamma h'' - \rho g h + f_{DEP} = P. \qquad [\text{S2.2}]$$

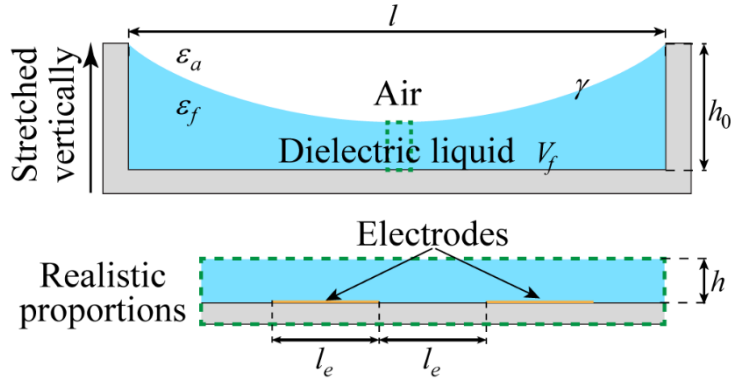

**Fig. S2.** *Two-dimensional illustration of the two-electrodes configuration and the relevant physical parameters used in modeling the system. (A) A dielectric liquid of volume $V_f$ is placed in a chamber of length $l$ and height $h_0$, forming a thin film wetting the chamber's floor and walls. Two surface electrodes of width and gap $l_e$, are located at the center of the chamber. The dielectric permittivity of the fluid and air above it are $\varepsilon_f$ and $\varepsilon_a$, respectively, and the surface tension of the fluid-air interface is $\gamma$. (B) A closer view on the electrode region. Since the dimensions of the electrodes are significantly smaller than the size of the chamber, we assume an approximately constant height of the liquid film in this region for the purpose of electric field and force calculations.*

Using the following non-dimensional parameters, $h = h_0 \eta$, $x = \xi l / 2$, we obtain the non-dimensional equation for the interface shape,

$$\eta'' - \text{Bo}\,\eta + \frac{l^2}{4\gamma h_0} f_{DEP} = C_P, \qquad [\text{S2.3}]$$

where $\eta$ is the non-dimensional height of the fluid, $\xi$ is the non-dimensional spatial coordinate along the chamber, and $\mathrm{Bo} = \rho g l^2 / 4\gamma$ and $C_P$ are the Bond number and the non-dimensional pressure constant. As presented in Fig. 3A, the numerically obtained DEP force distributions strongly resemble a Gaussian distribution. To facilitate an explicit expression for the deformation, we thus approximate the $f_{DEP}$ as a Gaussian of width $l_e$, and an amplitude $a$ set such that its total force matches $F_{DEP}$,

$$f_{DEP} = a \exp\left(-\left(\frac{x}{l_e}\right)^2\right), \quad a = \frac{F_{DEP}}{l_e \sqrt{\pi}\, \mathrm{erf}\left[l / 2 l_e\right]}. \quad [\text{S2.4}]$$

Following the above assumption of Gaussian DEP force distribution, equation (S2.5) for the liquid shape under DEP actuation using pair of electrodes configuration takes the form,

$$\eta'' - \mathrm{Bo}\,\eta + \mathrm{DEP}\exp\left(-\frac{\xi^2}{c^2}\right) = C_P, \quad [\text{S2.5}]$$

where $\mathrm{DEP} = \dfrac{a l^2}{4 \gamma h_0}$ represents the non-dimensional amplitude of the force and $c = \dfrac{2 l_e}{l}$ represents the non-dimensional width of the Gaussian. We assume that the liquid is pinned at the edges of the chamber, $\eta(-1) = \eta(1) = 1$, providing two boundary conditions, and in addition require the total fluid volume to be conserved, $\int_{-1}^{1} \eta(\xi) d\xi = C_V$, providing the remaining conditions for resolving the pressure $C_P$.

The general solution for the shape of the interface, is given by,

$$\eta(\xi) = \frac{e^{-\sqrt{Bo}\,\xi}}{4 Bo \left(1 + e^{2\sqrt{Bo}}\right)} \left(4\left(Bo\, e^{\sqrt{Bo}} + e^{\sqrt{Bo}} C_P - e^{\sqrt{Bo}\,\xi} C_P - e^{\sqrt{Bo}(2+\xi)} C_P + e^{\sqrt{Bo}(1+2\xi)}(Bo + C_P)\right)\right.$$
$$- \sqrt{Bo}\, c\, \mathrm{DEP}\, e^{\frac{Bo c^2}{4}} \sqrt{\pi} \left[\left(1 + e^{2\sqrt{Bo}\,\xi}\right)\mathrm{Erf}\left[\frac{1}{c} - \frac{\sqrt{Bo}\,c}{2}\right] - \left(e^{2\sqrt{Bo}} + e^{2\sqrt{Bo}(1+\xi)}\right)\mathrm{Erf}\left[\frac{1}{c} + \frac{\sqrt{Bo}\,c}{2}\right] + \quad [\text{S2.6}]$$
$$\left. + \left(1 + e^{2\sqrt{Bo}}\right)\left(\mathrm{Erf}\left[\frac{\sqrt{Bo}\,c}{2} - \frac{\xi}{c}\right] + e^{2\sqrt{Bo}\,\xi}\mathrm{Erf}\left[\frac{\sqrt{Bo}\,c}{2} + \frac{\xi}{c}\right]\right)\right]\right)$$

where the non-dimensional pressure term is,

$$C_P = \frac{2 Bo\left(-1 + e^{2\sqrt{Bo}}\right) - Bo^{3/2}\left(1 + e^{2\sqrt{Bo}}\right) C_V}{2\left(1 + \sqrt{Bo} + \left(-1 + \sqrt{Bo}\right) e^{2\sqrt{Bo}}\right)} +$$
$$+ \frac{\sqrt{Bo}\, c\, \mathrm{DEP} \sqrt{\pi}\left(\left(1 + e^{2\sqrt{Bo}}\right)\mathrm{Erf}\left[\frac{1}{c}\right] - e^{\sqrt{Bo} + \frac{Bo c^2}{4}}\left(\mathrm{Erf}\left[\frac{1}{c} - \frac{\sqrt{Bo}\,c}{2}\right] + \mathrm{Erf}\left[\frac{1}{c} + \frac{\sqrt{Bo}\,c}{2}\right]\right)\right)}{2\left(1 + \sqrt{Bo} + \left(-1 + \sqrt{Bo}\right) e^{2\sqrt{Bo}}\right)}. \quad [\text{S2.7}]$$

To present a simplified solution we solve for the case of $Bo = 0$ and also assume an initially flat interface, i.e., $C_V = 2$, which simplifies the solution even further. The Bond number in our system is not negligible $Bo \approx 10$ yet, the solution for the deformation of the simplified case with initially flat interface,

$$d(\xi) = A\left(\left(-8e^{-\frac{\xi^2}{c^2}} + 2e^{-\frac{1}{c^2}}\left(1+3\xi^2\right)\right)c + \sqrt{\pi}\left(\left(2+6\xi^2-3c^2\left(\xi^2-1\right)\right)\text{erf}\left[\frac{1}{c}\right] - 8\xi\text{erf}\left[\frac{\xi}{c}\right]\right)\right), \quad \text{[S2.8]}$$

yielding decent approximation for the realistic case as presented in Fig. S3. Thus, because the Bond number, i.e., the gravity effect on the system, indeed alter the initial shape of the fluid. However, when examining the deformation, we subtract the initial shape from the shape of the interface after actuation. This way, in order to test the importance of gravity in the system, one needs to compare the Bond number with the DEP number while choosing an appropriate scaling to the length scale such as the width of the DEP force distribution.

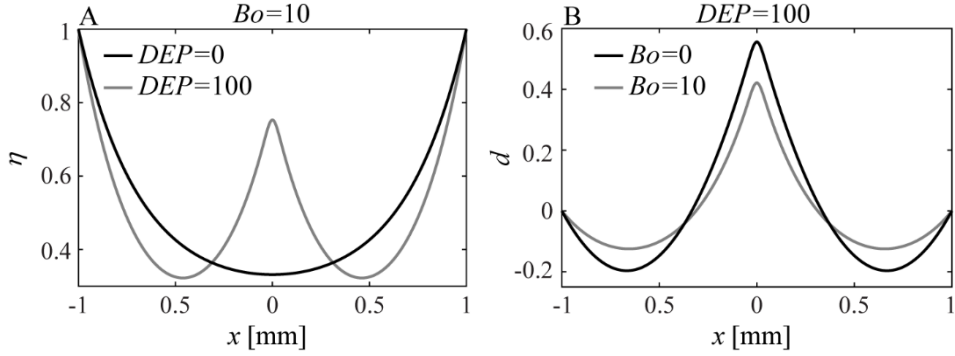

***Fig. S3.*** *Solutions for the DEP induced deformations of the interface using pair of electrode configuration with and without the effect of gravity for typical non-dimensional actuation and Bond number values. (A) The shape of the interface before (black solid line) and after (gray solid line) actuation for $Bo = 10$. (B) Comparision of the induced deformations with (gray solid line) and without (black solid line) gravity for $DEP = 100$. The black line represents the deformations of the interface for the initialy flat case, solution of equation.[S2.5], and the gray line represents the deformation the interface presented in A, i.e., substracting the initial shape of the interface from the inteface shape after actuation.*

## 3. DHM measurements interpretation

Digital holography microscopy (DHM) is a holography method which records a hologram image on a digital sensor, i.e., CCD or CMOS camera. The reconstruction of the image is done using numerical algorithms allowing fast acquisition and reconstruction of holograms in real-time. In this work we used the Lyncee Tec R1003 is a DHM working in reflection mode. In this mode the phase shift of the wave reflected from the measured surface is reconstructed using the hologram image.

In our experimental setup the liquid film is very thin, thus when we try to focus on the liquid-air interface, the floor of the chamber (the electrodes surface) is still in the coherence length of the microscope (the coherence length of the DHM is $200\,\mu m$). Thus, data gathered from the liquid-air interface contains interference pattern resulting from the floor. Moreover, the bottom surface consists of areas which are only glass, and areas of glass covered with $6\,nm$ of metal (the electrodes) which creates distortions of the measurements above the electrodes. To overcome this issue, we work in "semi-reflective mode", in which we focus on the very last surface in our device, the back side of the glass, and by the phase shift measurements obtained from this surface we calculate the surface topography.

Figure S4 presents schematic illustration of the two working modes. In the regular reflection mode, we can write the phase shift of the two beams and the difference between them as,

$$\varphi_1 = 2\frac{2\pi}{\lambda}h_1 n_a, \quad \varphi_2 = 2\frac{2\pi}{\lambda}h_2 n_a$$
$$\Delta\varphi = \varphi_2 - \varphi_1 = 2\frac{2\pi}{\lambda}n_a(h_2 - h_1) = 2\frac{2\pi}{\lambda}n_a \Delta h \quad \text{[S3.1]}$$

where $h$ and $\varphi$ represents the distance of the surface from the objective and the phase shift of the reflected beam respectively, $\lambda$ is the wavelength of the laser beam and $n_a$ is the refractive index of the air. To calculate the surface topography, we multiply the phase shift data obtain from the DHM by the following conversion factor,

$$\Delta h = \underbrace{\frac{\lambda}{4\pi n_a}}_{\text{Conversion factor}} \cdot \Delta\varphi. \quad \text{[S3.2]}$$

This conversion factor is one of the data the DHM provides when measuring a surface topography. However, if one focus on the lower surface as explained above, this factor needs to be modified. Based on the same process we did for the reflective mode we write the phase shift and phase difference of the two beams except for the location of the reference surface, which is now the bottom surface,

$$\varphi_1 = 2\frac{2\pi}{\lambda}h_1 n_a + 2\frac{2\pi}{\lambda}t n_f, \quad \varphi_2 = 2\frac{2\pi}{\lambda}h_2 n_a + 2\frac{2\pi}{\lambda}(t-\Delta h)n_f$$
$$\Delta\varphi = \frac{4\pi}{\lambda}\left(h_2 n_a - \Delta h n_f - h_1 n_a\right) = \frac{4\pi(n_a - n_f)}{\lambda}\Delta h \quad \text{[S3.3]}$$

where $t$ and $n_f$ are the smallest thickness and the refractive index of the transparent matter as illustrated in Figure 4SB. We note that although the beams reach the bottom surface, the thickness of the transparent matter is, $t$, cancel out from the equation. For that reason, adding more transparent layers with uniform thickness does not alter the conversion factor of our semi-reflective configuration,

$$\Delta h = \underbrace{\frac{\lambda}{4\pi(n_a - n_f)}}_{\text{Conversion factor}} \cdot \Delta\varphi \qquad [\text{S3.4}]$$

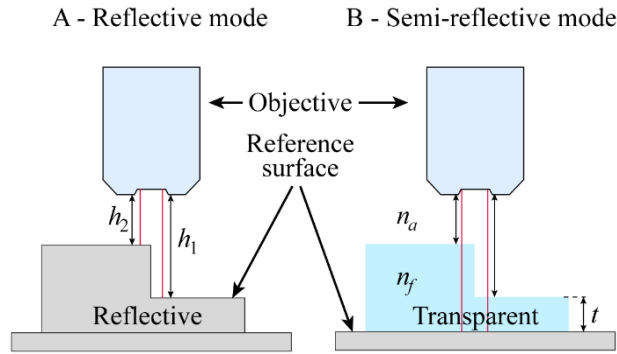

*Fig. S4.* Schematic illustration of the reflective and the semi reflective modes measuring the exact same topography made of reflective and transparent matters. The distance between the objective and the surface are $h_1$ and $h_2$ for the farther and closer surface to the objective respectively and $t$ is the distance thickness of the smaller step. $n_a$ and $n_f$ are the refractive indices of the transparent matter and the air, respectively. (A) present the normal working mode of the DHM where we focus on the surface we wish to measure and (B) presents the Semi-reflective mode where the focus is on different surface located below the interface we wish to measure.

## 4. Surface Roughness measurements using the digital holographic microscope

We measure the surface topography of an atomically polished silicon wafer with surface roughness of sub nanometer. By examine a field of view of $50 \times 50 \mu m^2$ (containing 86X86 data points), we yield a surface roughness with an RMS value of $3.7\,\text{nm}$ using the digital holography microscope. Thus, showing the limit of the DHM as a measurement tool for surface quality.

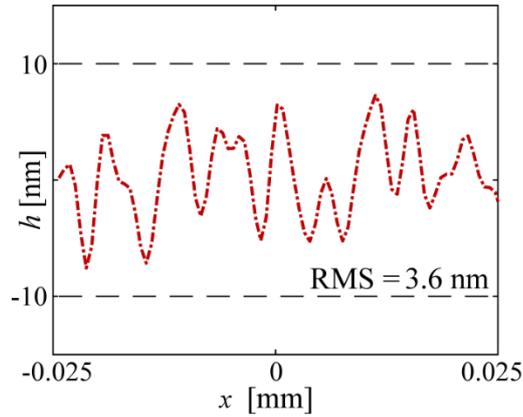

***Fig. S5.*** *Typical cross section measurement ($50\,\mu m$ in length) of an atomically polished wafer for estimation of the surface quality measurements capabilities using the DHM, yielding an RMS value of $3.6\,\text{nm}$.*

## 6. SI references

## Author Contributions

I.G., A.G., and M.B. conceived the research; I.G. performed the experiments, led the development of the models, and analyzed the data; F.P fabricated the devices and contributed to the design of the experiments; E.B. and A.R. contributed to the model development; A.G., and M.B. directed the research; I.G., F.P., E.B., A.G., and M.B. wrote the paper.